RESEARCH ARTICLE | AUGUST 30 2024

# Integrated magneto-photonic non-volatile multi-bit memory

H. Pezeshki  ; P. Li  ; R. Lavrijsen  ; M. Heck  ; B. Koopmans

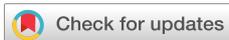

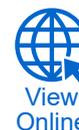

View
Online

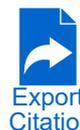

Export
Citation

# Integrated magneto-photonic non-volatile multi-bit memory

View Online     Export Citation     CrossMark

H. Pezeshki,[1,2,a)] 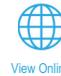 P. Li,[1] 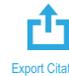 R. Lavrijsen,[1,2] M. Heck,[2] and B. Koopmans[1,2]

## AFFILIATIONS

[1]Department of Applied Physics and Science Education, Eindhoven University of Technology, Eindhoven 5612 AZ, The Netherlands
[2]Eindhoven Hendrik Casimir Institute, Center for Photonic Integration, Eindhoven University of Technology, Eindhoven 5600 MB, The Netherlands

[a)]**Also at:** EFFECT Photonics B.V., High Tech Campus, HTC 37, 5656 AE Eindhoven, The Netherlands.
**Author to whom correspondence should be addressed:** h.pezeshki@tue.nl

## ABSTRACT

We present an integrated magneto-photonic device for all-optical switching of non-volatile multi-bit spintronic memory. The bits are based on stand-alone magneto-tunnel junctions, which are perpendicularly magnetized with all-optically switchable free layers, coupled onto photonic crystal nanobeam cavities on an indium phosphide based platform. This device enables switching of the magnetization state of the bits by locally increasing the power absorption of light at resonance with the cavity. We design an add/drop network of cavities to grant random access to multiple bits via a wavelength-division multiplexing scheme. Based on a three-dimensional finite-difference time-domain method, we numerically illustrate a compact device capable of switching and accessing at least eight bits in different cavities with a 5 nm wavelength spacing in the conventional (C) telecommunication band. Our multi-bit device holds promise as a new paradigm for developing an ultrafast photonically addressable spintronic memory and may also empower novel opportunities for photonically driven spintronic-based neuromorphic computing.

## I. INTRODUCTION

Nowadays, there is no doubt on the critical role of integrated photonics in telecommunications for achieving ultrafast, and energy-efficient on-chip and inter-chip data transmission.[1–3] Moreover, there has been tremendous progress in novel application perspectives, such as in optical neural network computing and quantum information processing.[4–6] As for the power and delay overhead related to data transportation,[5,7,8] encoding information as optical bit patterns in the photonic domain without any opto-electronic conversion steps can be beneficial from the perspectives of modulation speed, power consumption, data transfer bandwidth, and cross talk. An important aspect of integrated photonics to extend its application perspectives relies on the implementation of photonic memories. Using these, versatile functionalities in signal processing, such as buffering and filtering,[9,10] optical computing,[11,12] data storage,[13,14] and quantum information,[15] can be realized.

Photonic memories are typically implemented in several ways. For optical buffers in signal processing,[9,10] microring resonators and delay lines are used to temporarily store the optical signals before their degradation. In view of the very large variation in the rate of incoming data, excessive area and energy overhead are required. Recently, non-volatile memories, such as phase-change materials (PCMs) and nonlinear optical materials, whose optical and electrical properties change upon optical and electrical stimuli,[16–21] are being considered an alternative due to their much longer retention time, lower power consumption, higher optical contrast, and fast switching speed.[4,6,20,21] Nonetheless, PCM-based memories would require large transient power and slow writing time in the range of nanoseconds for switching only a single bit

12 October 2024 11:14:55

due to the intrinsic atomic dynamics.[14,22] Addressing multiple bits using these memories would lead to devices with very large footprints[23,24] due to the use of several microring resonators.

It has been envisioned that an attractive alternative could be derived from recent breakthroughs, demonstrating that some materials platforms based on spintronics can be all-optically switched (AOS) and read (AOR).[25–27] The characteristic differences are rooted in the ultrafast spin dynamics[28] whereby the optical switch can be completed within picoseconds with high energy efficiency[29] compared to relatively slow phase-change processes. Introducing spintronic material based devices would not only enlarge the library of photonic memories, but also extend the functionalities of the entire platform by hybridizing it with integrated spintronics. Such a concept has comprehensively been investigated based on both alloys and the layered cobalt and gadolinium (Co and Gd) material system, showcasing AOS using optical pulses with incident energies in the range of femtojoule (fJ).[30–33]

Recently, Becker *et al.*[34] proposed a building block for spintronic–photonic integration for accessing magneto-tunnel junctions (MTJs) by vertically projecting light onto them using a two-dimensional grating coupler. However, such an approach comes with a large footprint and diffraction limit drawbacks. Then, Pezeshki *et al.*[35,36] proposed a hybrid plasmonic-photonic device for integrating spintronic racetrack memory with a photonic waveguide to access nanoscale ferrimagnet bits by injecting fJ (sub) picosecond light pulses through the waveguide. Nonetheless, as Kimel and Li[29] stated, incorporating magneto-photonic devices, such as opto-MTJs[37] in MRAM data storage applications, necessitates a large-scale integration of MTJ building blocks, randomly accessible at will for AOS and AOR. Kimel and Li addressed the issue of selective accessibility by proposing time-division and wavelength-division multiplexing schemes (TDM and WDM).[29] The TDM is a time-domain optical approach in telecommunications, where a number of optical signals are periodically interleaved in time, transmitted together, and then separated according to different arrival times. In contrast, the WDM technique is a frequency (wavelength)-domain optical approach in which optical signals with different wavelengths are combined, transmitted, and later separated based on their wavelengths. They suggested a method exploiting a cross-bar array of optical waveguide-based Bragg grating resonators for the purpose of TDM and WDM.[29] Although an interesting proposal, their proposed far-field method comes with a large footprint, low out-coupling efficiency, inefficacy from the power consumption aspect, and according to the authors, their WDM scheme requires many lasers at different wavelengths.

In this paper, we propose and numerically explore energy-efficient device functioning based on a near-field WDM approach through the use of a compact add/drop network[38] of photonic crystal (PhC)[39,40] nanobeam cavities on the indium phosphide (InP) membrane on the silicon (IMOS) platform. In our device, each MTJ, with a diameter of 60 nm acts as a 1 bit, is coupled onto the center of an individual PhC cavity with a distinct resonance wavelength. Using a three-dimensional finite-difference time-domain (FDTD) method, Lumerical FDTD,[41] we numerically demonstrate that by changing the wavelength of incident light, our device can enable wavelength-dependent switching of at least eight bits, located at different cavities with 5 nm wavelength spacing. The device concept is generic and can be implemented in other major photonic platforms, such as silicon photonics (SiPh)[42,43] and silicon nitride (SiN)[44,45] photonics. Our proposed approach makes a step forward toward spintronic–photonic integration, which can be used as a building block in a low-loss network of cascaded photonic interference couplers[46] to empower AOS of larger data streams, providing with ways to energy-efficient and fast access to spatially separated MRAM devices.

## II. DEVICE DESIGN

### A. Device layout

In this work, we propose a device design for multi-bit spintronic memory based on an add/drop network, which is shown schematically in Fig. 1. As sketched, it involves a photonic bus line, laterally placed between two photonic drop lines. The bus line is a meandering InP waveguide for carrying optical incident light within the network. The drop lines consist of InP nanobeam PhC cavities, i.e., grating-based cavities, $C_1$–$C_8$, each of which have one resonance wavelength and resonating at a particular wavelength, required for our WDM scheme. A cylindrical MTJ is evanescently coupled at the center of each cavity with a unique resonance wavelength (in Fig. 1 represented by a distinct color). Each MTJ consists of two ferromagnetic layers of a top pinned reference layer and an AOS-enabled free layer stacked to each other via a non-magnetic insulating metal oxide tunneling barrier (refer to Wang *et al.*[37] for more information). The free layer in our work is considered based on the Co/Gd material system, a (3d) transition metal, and a (4f) rare-earth ferromagnet essential for AOS,[29,30,47] whose threshold fluence has been priorly characterized.[33,48] The device works with in-coupling a picosecond optical Gaussian pulse into a fundamental transverse electric ($TE_0$) mode of the bus waveguide. As the $TE_0$ mode propagates through the waveguide, depending on the wavelength of the incident pulse, the pulse energy is transferred only to the designated cavity upon fulfillment of resonance conditions. Upon resonance in a cavity, the localized field enhancement by the cavity increases the amount of power absorbed in the coupled MTJ, leading to its ultrafast toggle-AOS upon fulfilling the required fluence threshold.[31,32,36,48] Following the subsequent switching of the free layer (FL) (see a conceptual diagram of a MTJ in the inset of Fig. 1), the magnetization of the reference layer (RL) is not perturbed by the laser pulse as a result of RKKY coupling with another hard magnetic layer (HL) and much higher Curie temperature.[37,49] The switching event can then be sensed electrically by the change in tunneling magnetoresistance (TMR) in the MTJ. It is worth mentioning that MTJs are electrical devices requiring bottom and top electrodes, where the bottom electrodes are very thin films, less than 10 nm, of transparent indium tin oxide[37,50] in order to allow the light passage. This thin film contributes negligibly to the design and performance characteristics but complicates the modeling and discussion. Therefore, for the sake of simplicity without losing generality, these electrodes are neglected, inspired by the work of Chen *et al.*[50]

In order to design a multi-bit device using the proposed WDM scheme, there are two important design steps that need to be considered for having a desired device performance, which will be further elaborated on in the rest of this section.

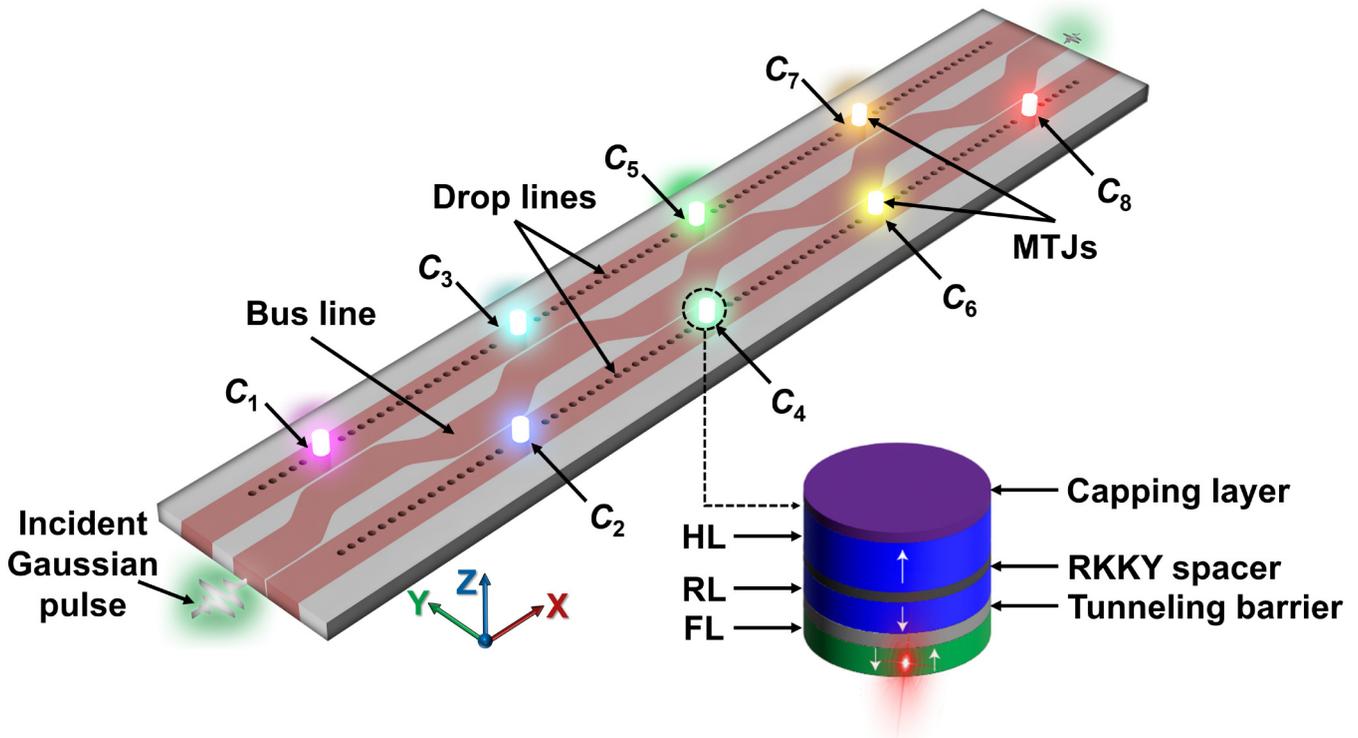

**FIG. 1.** Conceptual illustration of the device design as a building block for multi-bit spintronic memory. The device is an add/drop network of eight spatially separated nano-beam photonic crystal (PhC) cavities to randomly access magnetic bits presented by optomagnetic-tunnel junctions (MTJs). The cavities are denoted as $C_1$–$C_8$. Different colors enlightening the cavities showcase the monotonic change of the resonance wavelength for illustration purposes, which should not be confused with the realistic wavelengths within the telecommunication C-band, which are separated by 5 nm in this work. The inset shows a schematic diagram of the MTJ. HL, RL, and FL stand for the hard layer, the reference layer, and the free layer, respectively. The RKKY spacer provides an indirect coupling mechanism known as Ruderman–Kittel–Kasuya–Yosida (RKKY) coupling, in which the magnetic coupling between spins is indirectly mediated by conduction electrons. The dimensions are not in scale for the sake of clarity.

## B. Design criteria

We consider an add/drop network of eight cavities tuned to various wavelengths across the telecommunication C-band to all-optically switch an MTJ depending on the wavelength of the incident light. The challenge is to switch each bit independently without unintended switching of any other MTJs. We first need to provide enough energy for switching the entire bit, which is called the saturation energy $U_{sat}$. We need to ensure that the minimum pulse energy, $U_{min}$, fed to the device for successful switching of the targeted bit in the cavity $C_j$ tuned to its resonance wavelength, $\lambda_j$, must be equal or above the saturation energy, i.e., $U_{min}[j] \geq U_{sat}[j]$. The pulse energy may off-resonantly be coupled to undesignated cavities due to cross talk. It is then equally important to guarantee that the switching will not occur at any other cavities $k \neq j$ for each $\lambda_j$ due to cross talk. In other words, the amount of the pulse energy in any unwanted cavity should be kept less than the threshold energy, $U_{th}$, for which at least a fraction of the bit is switched. Thus, the maximum energy limit for switching the bit $j$, $U_{max}[j]$, without unwanted switching of other bits $k$ should be less than

$U_{th}[k]$. Hence, the goal is to find the range of the incident pulse energy fulfilling the above-mentioned two criteria. Without loss of generality, we suppose that the full-width at half-maximum (FWHM) of each resonance cavity ($\Gamma$), the wavelength spacing between each subsequent cavities, and the total absorption loss by each cavity ($L$) and MTJs ($A$) are identical. Therefore, saturation, $U_{sat}$, and threshold energies, $U_{th}$, of MTJs will be identical.

The absorption loss profile of the cavities is obtained by fitting based on a Lorentzian function as follows:

$$L(\lambda_0, \lambda_r, \Gamma) = \frac{L_0 \Gamma^2}{(\lambda_0 - \lambda_r)^2 + \Gamma^2},$$  (1)

where $L_0$ is the absorption loss by the cavity at resonance and $\lambda_0$ and $\lambda_r$ are the incident and resonance wavelengths, respectively. According to the first criterion and considering cross talk contributed by cavities upstream the network, the minimum pulse energy for successfully switching the bit $j$, $U_{min}[j]$, can be obtained from

$$U_{min}[j]\left(\prod_{i=1}^{j-1}(1-L[i,j])\right)L[j,j]A[j,j] \geq U_{sat}[j] \quad \forall j = 1, \ldots, 8, \tag{2}$$

where $L[i,j]$ is the absorption loss contributed by the upstream cavity $i$ when the incident wavelength is tuned to the resonance wavelength of the $C_j$ cavity, i.e., $\lambda_j$. The $L[j,j]$ and $A[j,j]$ are the $C_j$ cavity's absorption loss and the fraction of the absorption loss in that cavity, which is used to perform AOS on the target bit $j$, respectively. Note that the reflection off the designed cavity is so small that the possible interference between reflected light off the cavity and light in the bus waveguide can be neglected. Based on the second criterion, the incident pulse energy at $\lambda_j$ should be low enough such that the energy deposition in the $C_k$ cavity is lower than the threshold energy $U_{th}[k]$, required for switching at least a part of the area of any other bit $k \neq j$. Thus, there is an upper limit for the incident pulse energy, $U_{max}[j,k]$, which can be calculated as follows:

$$U_{max}[j,k]\left(\prod_{i=1}^{k-1}(1-L[i,j])\right)L[k,j]A[k,j] < U_{th}[k],$$
$$\forall k \neq j = 1, \ldots, 8, \tag{3}$$

where $L[k,j]$ and $A[k,j]$ are the $C_k$ cavity's absorption loss and the fraction of the absorption loss in that cavity for the incident pulse energy at $\lambda_j$, respectively.

## C. Implementation of design criteria

In order to demonstrate the design criteria, we emphasize more specifically the two essential parameters of a resonance cavity, that is the resonance width, $\Gamma$, and efficiency. We apply the above-mentioned concepts to analytical energy absorption analysis of four typical cases: a non-working device, a conditionally working device, a working device, and a device with less efficient resonance cavities. To do so, as a concrete example, we consider $L_0 = 0.5$, which is an initial absorption loss of 50% by cavities at resonance, $A = 1$, that is 100% of the loss being absorbed by the magnetic bit, $U_{sat} = 1$ as a normalization, and $U_{th} = 0.9$, i.e., 90% of $U_{sat}$, motivated by the fact that the light absorption is non-uniform in an MTJ (see Pezeshki et al.[36] for more insights). We further chose the incident wavelengths from the C-band with 5 nm spacing. Note that as of here, we will specify "threshold" pulse energies as the fraction of $U_{sat}$ in dimensionless units.

### 1. Non-working case

We first argue the decisive role of the narrow resonance width of a cavity by assuming a non-working case, for which we consider cavities with a broad resonance spectrum, for instance, with $\Gamma = 30$ nm. We plotted $U_{min}$ and $U_{max}$ as defined in Eqs. (2) and (3) in Fig. 2(a) for this case. In this figure, the minimum energy required to switch the targeted bit in the cavity $C_j$ at $\lambda_r[j]$ for $j = 1$ to 8, $U_{min}$, is indicated by the solid purple line. The dashed and dotted lines represent the upper limit of the energy required for switching the targeted bit in $C_j$, $U_{max}[j,k]$, without unwanted switching of all other bits $k \neq j$, i.e., $U_{max}[j,k] > U_{min}[j]$ for all $k \neq j$. From this point onward, we neglect the index $[k]$ for $U_{max}$; e.g., $U_{max}[j = 1, k]$ is denoted as $U_{max}[1]$ for simplicity. According to Fig. 2(a), by going downstream, $U_{min}$ increases because of the absorption loss in the earlier cavity/cavities in the chain. Based on this plot, at the $C_1$ cavity, when using its resonance wavelength for switching the bit, very high energies are required to switch other bits in the $C_2$–$C_8$ cavities, as they are off-resonant and also further downstream. We need higher energy for switching the targeted bit at the $C_8$ cavity, i.e., $> 43$, which exceeds the essential energy level for switching the bits in other cavities, i.e., $\leq 20$, as highlighted by a red-dotted ellipse in Fig. 2(a). In this case, by switching the bit in the $C_8$ cavity, other bits further upstream get switched-on erroneously by the energy leakage due to cross talk. Thus, in this case, at least one bit will be inadvertently switched together with the targeted bit.

### 2. Conditionally working case

For the second case, by reducing the resonance width from $\Gamma = 30$ nm to $\Gamma = 6$ nm, we can have a device that conditionally works. The result for this case is shown in Fig. 2(b). In contrast to the case in Fig. 2(a), the $U_{min}$ limit required to switch the target bit in the cavity $C_j$ at $\lambda_r[j]$ for $j = 1$ to 8 is below the $U_{max}$ limit needed for unwanted switching of the bits $k \neq j$. Even though $U_{min} < U_{max}$, there is only a marginal difference for cavity numbers $j > 1$. In addition, $U_{min}[j = 8] > U_{max}[j = 2, k = 1]$ and $U_{max}[j = 8]$ as highlighted by a red-dotted ellipse. This means that, for instance, by applying a fixed input pulse energy of $U_{min}[j = 8]$ for switching the bits in the $C_2$ and $C_3$ cavities, we suffer from unwanted switching of the bits in the $C_1$ and $C_2$ cavities, respectively. In other words, the device can only be made to work by choosing dynamically lower pulse energy for switching the targeted bits upstream. We need a higher energy to switch the bit in the $C_8$ cavity (without unwanted switching elsewhere upstream) and a lower one when targeting the bit in the $C_2$ and $C_3$ ones. This requirement will lead to a less trivial operational scheme in the sense that laser energy must be tuned carefully at each different wavelength.

### 3. Unconditionally working case

Our analysis shows that the cavities with a resonance width of $\Gamma < 4$ nm make the device work unconditionally. We introduce such a working device by choosing $\Gamma = 3$ nm, where the result is shown in Fig. 2(c). Based on this plot, the $U_{min}$ limit is well below all the $U_{max}$ limits, i.e., less than the threshold energies of all other bits in the off-resonant cavities. For a pulse energy between ~3 and ~7 as depicted by a red double-headed arrow, we only have events of desired switching of the targeted bits, without falsely switching of other bits. Overall, there is a significant energy margin between the $U_{max}$ and $U_{min}$, which makes it possible to address all the bits in the cavities using the same laser energy, $U_{min}[j] = $ const. for all $j = 1-8$.

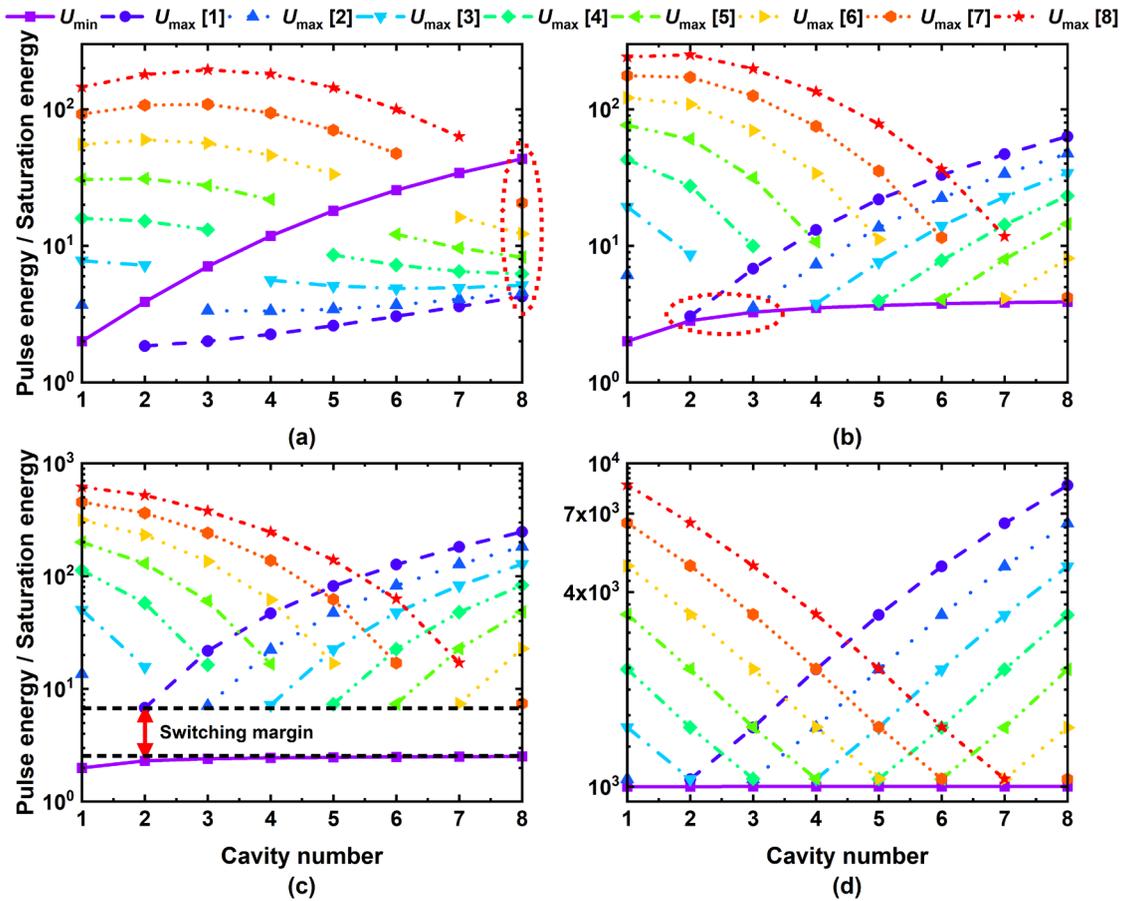

**FIG. 2.** Design considerations. Implementing the design criteria for four major cases of (a) a non-working device with a full-width at half-maximum (FWHM) of $\Gamma = 30$ nm, (b) a device with $\Gamma = 6$ nm that conditionally works, (c) a working device with $\Gamma = 3$ nm, and, finally, (d) a device with less efficient resonance cavities with $L_0 = 0.001$ for $\Gamma = 12$ nm. The $L_0$ in (a)–(c) is 0.5. Note that for the sake of simplicity, we neglected the index [$k$] for $U_{max}$; e.g., $U_{max}[j = 1, k]$ is $U_{max}$ [1].

### 4. Less efficient cavity case

Finally, for the last case, we demonstrate the role of the efficiency of the cavity by considering a device with much less efficient resonance cavities. To model such a device, we set $L_0 = 0.001$, 0.1% of the incident power is coupled to each cavity at its resonance. We show the result of an extreme case in which $\Gamma = 12$ nm [see Fig. 2(d)]. Based on this plot, we can see $U_{min}$ for switching downstream and upstream bits is almost equal and is below all the $U_{max}$ limits because of poor coupling condition between the bus and drop lines in the add/drop network. More importantly, this case unravels the inverse relation between the energy efficiency of switching and performance of resonance cavities. Here, correct switching of the bits in cavities with much wider $\Gamma$ can become possible but at the expense of much higher pulse energies compared to the conditionally working and working cases. Altogether, the results of this analytical study serve as a general guideline for designing our multi-bit spintronic memory.

## III. RESULTS

To address eight magnetic bits across the C-band using our proposed add/drop scheme, we should consider a case of a realistic device. For this purpose, we used a cavity that is formed via removing three holes across the axis of a 1D PhC lattice, known as an L3 cavity.[51] Figure 3(a) shows a schematic of this cavity loaded with an MTJ possessing a radius of $r_m = 30$ nm [see Fig. 3(b)], where the waveguides' width and height are $w = 570$ nm and $h = 280$ nm, respectively, and the gap between the bus line and drop lines around the coupling area is $g = 200$ nm. We designed and optimized eight different PhC cavities, with a lattice constant and a hole radius of $a = 370$ nm and $r = 85$ nm [see Fig. 3(b)], without the coupled MTJs to resonate across the wavelength range of 1535–1570 nm, considering the wavelength spacing of 5 nm achieved by modifying the size, $2r_d$, and/or position, $x_d$, of the point defects in purple at the right and left sides of the cavity [see Fig. 3(b)]. Table I shows the cavities' design parameters and resonance properties based on their normalized electric field resonance

12 October 2024 11:14:55

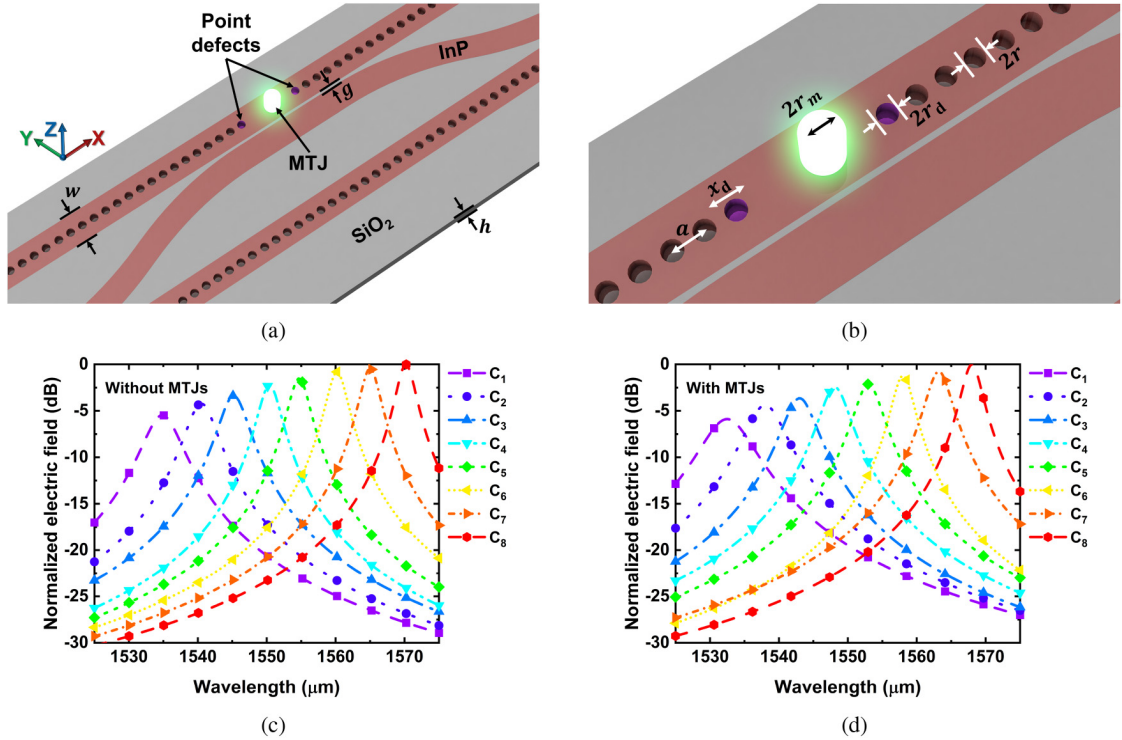

**FIG. 3.** Design and response of the cavities for the purpose of multi-bit spintronic memory. (a) A generic schematic of a L3 cavity loaded with an MTJ, where the wave-guide's width and height are $w = 570$ nm and $h = 280$ nm, and the gap between the bus line and drop lines around the coupling area is $g = 200$ nm. (b) The magnified view of the L3 cavity in which the PhC cavity's lattice constant and holes radii are $a = 370$ nm and $r = 85$ nm, respectively. The point defects of the cavity are shown in purple, where their radius and the spatial shift in their position are denoted by $r_d$ and $x_d$. The coupled MTJ is also shown in greenish glow color with the radius of $r_m = 30$ nm. (c) and (d) Normalized electric field resonance spectra of the designed cavities probed in the middle of the waveguide and at the center of the cavity for the two cases of without and with the coupled MTJs.

**TABLE I.** Design parameters and resonance properties of the cavities for the use in the add/drop network.

| Cavity number | $r_d$[a] (nm) | $x_d$[a] (nm) | $\lambda_{ru}$[b] (nm) | $\Gamma_{ru}$[b] (nm) | $\lambda_{rl}$[b] (nm) | $\Gamma_{rl}$[b] (nm) |
|---|---|---|---|---|---|---|
| $C_1$ | 103.5 | 18.5 | 1535 | 4.90 | 1533 | 7.50 |
| $C_2$ | 103.5 | 29.6 | 1540 | 4.00 | 1538 | 6.00 |
| $C_3$ | 103.5 | 38.5 | 1545 | 3.50 | 1543 | 5.25 |
| $C_4$ | 85 | 27.4 | 1550 | 2.90 | 1548 | 4.65 |
| $C_5$ | 85 | 34.8 | 1555 | 2.75 | 1553 | 4.20 |
| $C_6$ | 66.5 | 34.8 | 1560 | 2.50 | 1558 | 3.60 |
| $C_7$ | 61 | 46.3 | 1565 | 2.50 | 1563 | 3.58 |
| $C_8$ | 66.5 | 27.4 | 1570 | 2.45 | 1568 | 3.23 |

[a]The parameters $r_d$ and $x_d$ stand for the size of the point defects and the spatial shift in their position.
[b]$\lambda_{ru}$ (rl) and $\Gamma_{ru}$ (l) indicate the resonance wavelength and the full-width at half-maximum of the corresponding cavities in two cases of unloaded and loaded, respectively.

spectra, probed in the middle of the waveguide and at the center of the cavity, as shown in Figs. 3(c) and 3(d) at two conditions of without (unloaded) and with (loaded) the coupled MTJs. This table shows that the $\Gamma$ of unloaded cavities varies between 2.45 nm at $\lambda_r = 1570$ nm and 4.9 nm at $\lambda_r = 1535$ nm. On the contrary, the loaded cavities have slightly broader resonance spectra, $\Gamma = 3.23$ nm at $\lambda_r = 1570$ nm and $\Gamma = 7.5$ nm at $\lambda_r = 1535$ nm, due to the absorption losses imposed by the coupled MTJs.

To further illustrate the wavelength-selectivity of the device loaded with MTJs, the electric field distribution across the section surrounding the $C_4$ cavity at its resonance wavelength $\lambda_{rl} = 1548$ nm as well as the resonance wavelengths of its neighboring cavities $C_3$ and $C_5$, i.e., 1543 and 1553 nm, are presented in Fig. 4. Based on Figs. 4(a) and 4(c), at $\lambda_0 = 1543$ and 1553 nm, the coupling to $C_4$ is very weak and the incident light mostly propagates through the waveguide. However, the light coupling to $C_4$ becomes intensified at $\lambda_0 = 1548$ nm as shown in Fig. 4(b), for which the cavity provides $> 3\times$ electric field enhancement relative to the off-resonant cases. It can be seen that by de-tuning the wavelength of the incident light by $\pm 5$ nm, we cannot excite $C_4$ due to the weak coupling between the bus and drop lines.

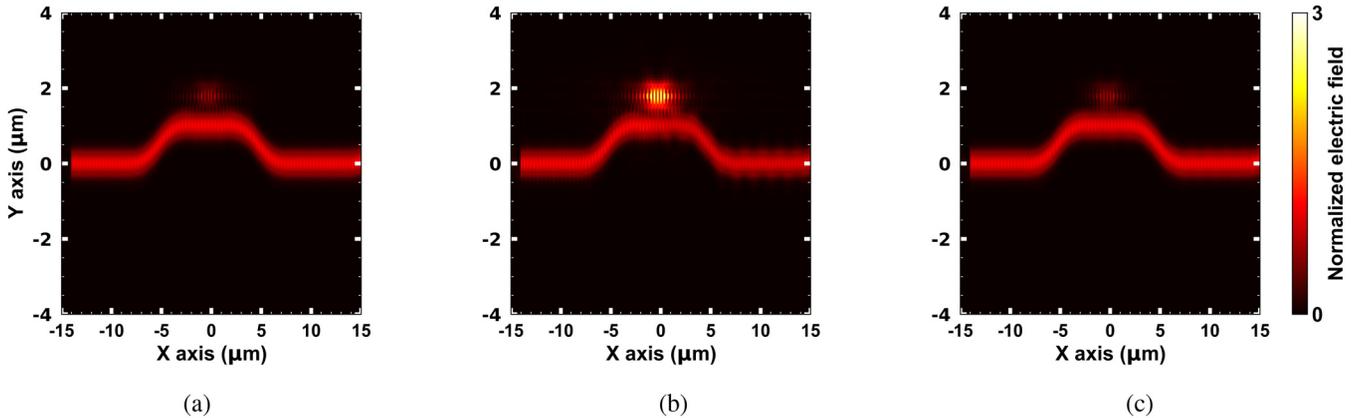

FIG. 4. Dependency of the cavity response on the wavelength. (a)–(c) Electric field distribution across the $C_4$ section for the incident light wavelengths of 1543, 1548, and 1553 nm, respectively.

All-optical switching of magnetization is induced by fast non-adiabatic heating[30,47] and a discrete, pulse length dependent threshold energy density. Hence, we evaluate the energy efficiency of AOS in our FDTD simulations based on the instant absorbed energy density in the coupled MTJs.[36] In our simulation, we extract the spatial distribution of the absorbed energy density by the magnetic switching layer, i.e., Co/Gd (represented by a Co layer instead in our simulation). Figure 5(a) shows the one-dimensional distribution of the absorbed energy density cut-through the middle of the Co layer of the MTJ in $C_4$ for the two devices of the bare waveguide and the proposed PhC device. The results are shown along the Y axis, which are integrated across the MTJ's surface along the X axis (refer to Fig. 3 for X and Y axes definitions) and normalized to the maximum absorbed energy. As shown, the $C_4$ cavity at its resonance, i.e., $\lambda_0 = 1548$ nm, shows 8× peak-to-peak enhancement in the absorbed energy density by the MTJ. In our previous work,[36] we theoretically showed that with 7× enhancement in the absorbed energy density, we can exceed the absorbed switching threshold of 0.5 mJ/cm² (Refs. 32 and 36) for an incident pulse energy of only 600 fJ without running into the regime of nonlinear absorption.

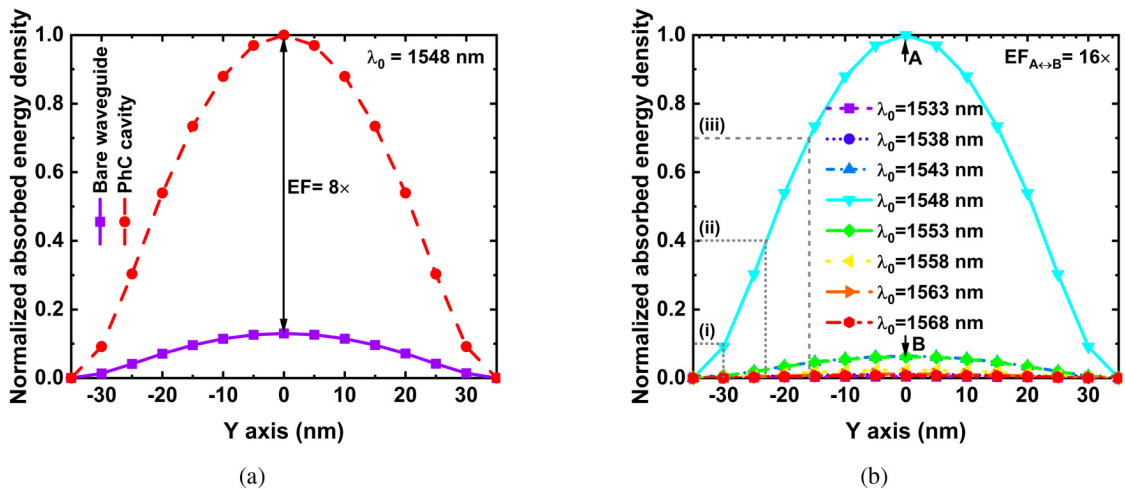

FIG. 5. Absorbed energy density distribution across MTJ. Normalized absorbed energy density along the Y axis cut-through the middle of the cobalt (Co) layer, integrated across the MTJ's surface in direction of the X axis (refer to Fig. 3 for X and Y axes definitions) for (a) a bare waveguide and the PhC cavity $C_4$ at $\lambda_0 = 1548$ nm and (b) the $C_4$ cavity for the incident light wavelengths from $\lambda_0 = 1533$ to $\lambda_0 = 1568$ nm, respectively. "EF" stands for the enhancement factor, and "A" and "B" point to the peak of the absorbed energy density curves at $\lambda_0 = 1548$ nm and $\lambda_0 = 1548 \pm 5$ nm, respectively. The three points indicated by (i), (ii), and (iii) indicate the cases of non-working, marginally working, and unconditionally working devices, respectively.

Therefore, it can be seen that the proposed cavity can provide the condition for AOS.

We have further illustrated the absorbed energy distribution across the MTJ at the $C_4$ as a function of the wavelength of incident light, from $\lambda_0 = 1533$ to $\lambda_0 = 1568$ nm, in Fig. 5(b). According to this figure, the absorbed energy density is maximum for $\lambda_0 = 1548$ nm, which is equal to the resonance wavelength of the loaded cavity $C_4$. Changing the wavelength of the incident light only by $\pm 5$ nm leads to a $16\times$ drop in the amount of energy absorbed density by the MTJ, i.e., from 1 to 0.0625. Note that since the absorption profile is not uniform across the MTJ, to be able to switch the whole MTJ, we need to increase the power of the incident light by ~90%, which may result in unwanted switching of a fraction of other bits in neighboring off-resonance cavities. Hence, it is crucial to conduct the study presented in Sec. II B to see if we can have a switching margin to guarantee the multiplexing performance of our device.

Based on the results of Figs. 5(b) and 3(d), we implemented the design criteria for three cases of non-working, marginally working, and unconditionally working devices, respectively, indicated by (i), (ii), and (iii) in Fig. 5(b). The analysis results are shown in Fig. 6. Note that in this analysis, the absorption loss by the loaded cavities was considered $L_0 = 0.25$ at their resonance condition based on the numerical FDTD simulations. For the non-working device, i.e., case (i) where $U_{th} = 0.1 \times U_{sat}$, which simply means that the maximum energy that we apply would be increased by 90% relative to the $U_{min}$, and the result in Fig. 6(a) indicates that very high energies are required to switch the bits in this case, which exceeds the required energy level for switching the bits in other cavities (see the red-dotted ellipse). In other words, due to the energy leakage, all other bits are switched-on inevitably in addition to the targeted bit. As for the case (ii) with $U_{th} = 0.4 \times U_{sat}$, in contrast to the conditionally working case in Sec. II B, the minimum energy $U_{min}$ is below the $U_{max}$ limit for all cavities $C_j$ for $j = 1$–8 [see Fig. 6(b)]. However, the differences between $U_{min}$ and the lowest of the $U_{max}$ for all cavities across the whole network are very marginal, which makes its multiplexing function susceptible to variations in the input pulse energy. That is why we called this case marginal. We then have the case where the device works unconditionally, i.e., the case (iii) with $U_{th} = 0.7 \times U_{sat}$, and the $U_{min}$ limit is well below all the $U_{max}$ limits across the network with a switching margin as large as 3.5 [see Fig. 6(c)]. In this case, similar to the unconditionally working case described in Sec. II B, we have all desired events of switching with a proper multiplexing function, thanks to its large switching margin.

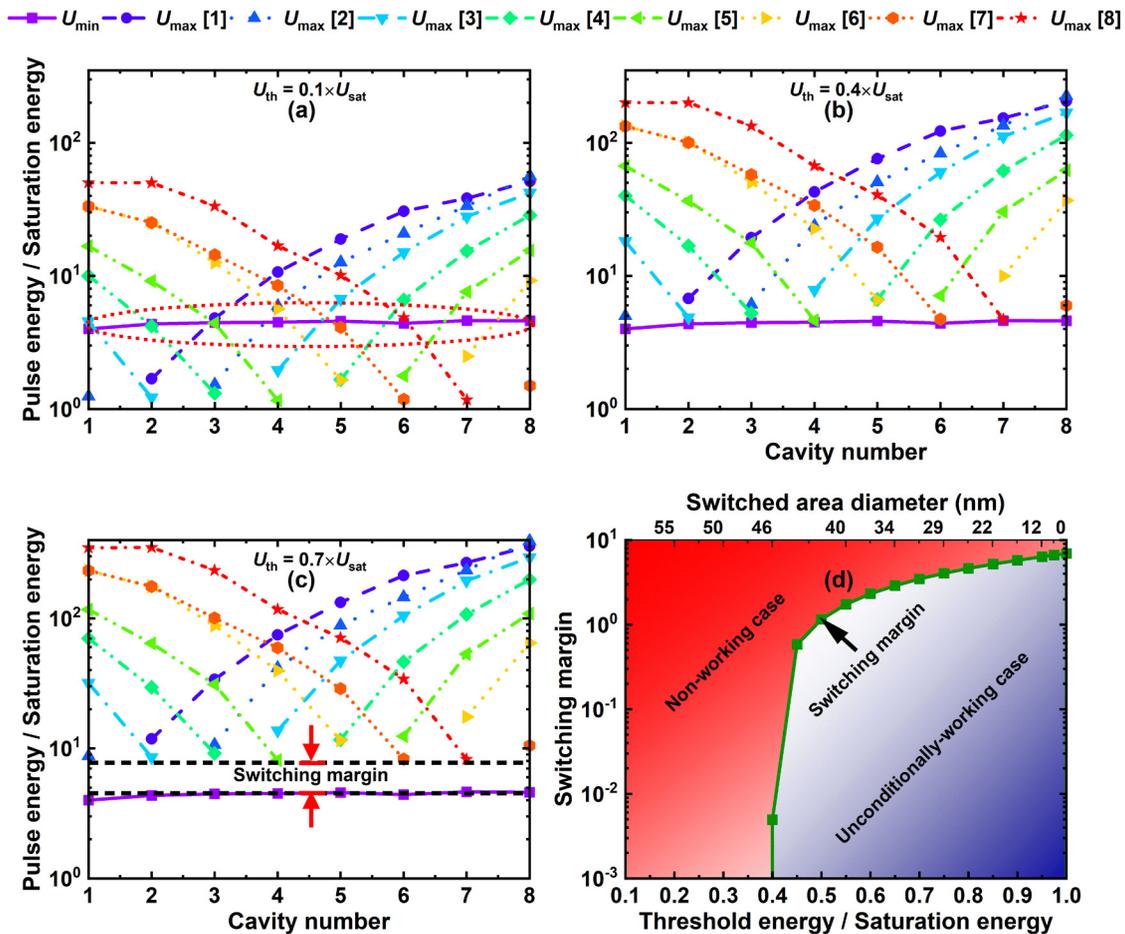

FIG. 6. All-optical switching (AOS) efficiency. Implementing the multiplexing criteria for the designed device in which the threshold energy $U_{th}$ is (a) $0.1\times$ the saturation energy $U_{sat}$, (b) $0.4 \times U_{sat}$, and finally, (c) $0.7 \times U_{sat}$. (d) Switching margin as a function of the switched area diameter in nm and the ratio of threshold energy to saturation energy. The $L_0$ for the designed cavity is 0.25 at resonance.

In two cases of marginally working and unconditionally working devices discussed in the previous paragraph, depending on the ratio of the threshold energy to the saturation energy, i.e., $U_{th}/U_{sat}$, an area of the targeted MTJ with a unique diameter can be toggle-switched. For instance, in the cases of $U_{th} = 0.4 \times U_{sat}$ and $U_{th} = 0.7 \times U_{sat}$, areas of the MTJ with diameters of 46 and 32 nm get switched-on, which are equal to 77% and 53.3% of the MTJ surface area, respectively [see the values on the X axis of Fig. 5(b) corresponding to the cases of (ii) and (iii)]. Therefore, one can see that there is a relation between the switching margin and both the $U_{th}/U_{sat}$ ratio and the switched area diameter. Therefore, we provided an overview plot in Fig. 6(d) showing the dependency of the switching margin, the green symbol-line curve, on the abovementioned parameters. According to this figure, for $U_{th}/U_{sat} \geq 0.4$, equivalent to switched area diameters of $\leq 46$ nm, the switching event with a proper multiplexing function can happen while its switching margin increases with an increase in the $U_{th}/U_{sat}$ ratio. It is worthwhile mentioning that considering the dynamics of domain walls and the size of the target MTJ, we can still have a single domain switching when the switched area is 53.3% of the MTJ surface area.

Overall, based on the combination of the numerical simulations and our proposed analytical study in Sec. II B, we illustrated a device concept to enable magneto-photonic non-volatile multi-bit spintronic memory based on the WDM scheme with at least eight bits in the telecommunication C-band.

## IV. DISCUSSION

Integrating opto-MTJs on photonic chips has been considered an interesting option for hybrid integration of spintronics and photonics from the switching speed and energy perspectives.[37,50,52] To enable the realization of this concept, it is advantageous to design a platform to provide large-scale integration of MTJs and parallel access for switching and reading at will. That is why in this work, we presented a device concept based on the near-field coupling WDM scheme in a PhC-based add/drop network, which is significantly advantageous in terms of the footprint and energy efficiency. In addition, our proposed device is completely monolithic, and its simplicity of integration could potentially allow for easy incorporation of spintronic building blocks. It is noteworthy mentioning that we designed our device for IMOS, but the concept is generic and can be seamlessly transferred to other integrated photonic platforms, most notably SiPh, which has similar waveguide geometries and material indices of refraction as IMOS, and with a lower two-photon absorption.

It is important to note the characteristics of AOS. Much higher energy above AOS threshold leads to a multi-domain state or thermal demagnetization.[31,53,54] There is a fluence gap between AOS and the multi-domain state. This fluence gap diminishes as pulse duration increases.[54] During the AOS event, it is essential that the part in which the absorption is the least should be switched. At the same time, for the same fluence energy, no multi-domain state should be triggered at the part where the absorption is the most. Consequently, it is necessary to guarantee that this absorption in-homogeneity should lie within the fluence window, $f_w$, which is defined as the ratio between the threshold fluence of the multi-domain state and AOS. As shown in Fig. 6, an in-homogeneity of $U_{th} = 0.4 \times U_{sat}$ is at the boundary to guarantee the functionalities of our designed building block. Thus, in this extreme case, the fluence window of $f_w > 2.5 (= 1/0.4)$ is at least necessary, imposing the longest pulse duration limit, reducing the effectiveness of a photonic crystal cavity. As the in-homogeneity decreases, the necessary fluence window is less. Fortunately, as discussed, smaller MTJ devices are inherently less susceptible to homogeneity discrepancies, thus necessitating a reduced fluence window. This coincidence is aptly exemplified in Fig. 6, ultimately enhancing likewise our switching margin. Based on the bandwidth requirement of our designed device discussed in Sec. III, we found a pulse duration equal to/longer than 1 ps is necessary, as the pulse bandwidth needs to fit into the resonator bandwidth. Recent experimental data[55] have demonstrated a fluence window of $f_w > 2$ at such a pulse duration, which validates our design scenarios as discussed in Figs. 6(b) and 6(c).

Our theoretical investigation detailed in Sec. II B also argued the benefit of decreasing the spectrum width of the cavity to increase the switching margin as well as to create the potential of fitting more cavities, i.e., more bits. Nevertheless, we note that care must be taken in this process. As the spectrum width narrows down, pulse duration needs to be increased accordingly, which reduces the tolerance for the above-mentioned multi-domain state. We, therefore, suggest to adopt AOS materials with longer maximum pulse duration, while at the operation pulse duration, the ratio between the threshold energy of thermal demagnetization and AOS needs to be enough substantial.

As explained in previous paragraphs, there are factors, such as the quality factor of the cavity, the absorption homogeneity of the MTJ and the pulse duration, as well as the material's AOS window influence the exact amount of bits that our device allows us to scale up to.[55] The rule of thumb is that the larger the switching margin is as well as the AOS switching window, the higher the number of bits that can be crunched within a single channel. Based on our estimations, if the diameter of the MTJ is reduced by a factor of two, the absorption in-homogeneity is reduced accordingly by a factor of two, therefore leading to twice the switching window. Thus, the amount of bits can be doubled, assuming a large AOS window. We estimated that this can be established without a major increase of required pulse energies. Moreover, there is a large engineering space available to potentially increase the number of bits further, but we consider this beyond the scope of our conceptual paper.

Finally, we address the energy efficiency and speed of our hybrid photonic–spintronic approach. So far, several memory concepts have been proposed, in which a photonic stimulus is used. There has been a very thorough review in comparison between different memristor technologies[56] and a comprehensive survey on phase-change memory (PCM) up to 2023.[57] Comparing these technologies based on the data presented in Refs. 56 and 57, the switching of an MTJ is in no doubt the fastest (ps switching[37]), which is owning to the ultrafast switching dynamics of AOS (see Ref. 29) as compared with other switching mechanisms, such as the slow dynamics of crystallization and glassification as in the case of PCM (>10 ns). In terms of switching energy, by scaling the devices of different technologies presented in Table I in Ref. 56 into the same area, our MTJ has a switching energy at least one order of magnitude lower than the lowest from Table I, PCM(O), in Ref. 56 according to Refs. 37, 49, and 58, with 30 pJ for $1\,\mu m^2$. More generally, our proposed technology offers the possibility of using the optical signals that enter the chip for writing the bits, that is, in fact, our ultimate vision. Here, we do not want to first electrically read out optical data to generate light pulses for writing the magnetic elements, but directly transfer information from the optical to the magnetic domain. This would avoid both the use of electro-optical detection as well as light generation by an additional laser. Therefore, we can possibly envision an off-chip memory, with the access time similar to SRAM and DRAM (using data from Refs. 59 and 60) leveraging the intrinsic advantages of our technologies.

## V. CONCLUSION

We have demonstrated an integrated hybrid photonic–spintronic device concept, enabling random AOS of eight nanoscale MTJs coupled onto an IMOS platform. The device comprises eight PhC nanobeam cavities with different resonance wavelengths covering the telecommunication C-band, within a bandwidth of 35 nm, in each of which an MTJ is coupled. The cavities offered significant spatially localized electric field enhancement thanks to the strong light–matter interaction, which leads to a substantial increase of the absorbed energy density in the coupled MTJs and consequent switching. We introduced an analytical approach as a general guideline for designing multi-bit spintronic memory based on add/drop networks working on the WDM scheme. Numerical FDTD results showed that our device can multiplex between the MTJs by de-tuning the wavelength of the incident light only by 5 nm, given the enhanced absorbed energy density in the coupled MTJs, thanks to PhC cavities. In this way, we can decisively and at the same time randomly switch any MTJ at will without unprecedented switching of any other MTJ in the network. We believe that this device offers good prospects as a building block for developing an opto-spintronic memory and photonic buffer technology with possible potential in newly-emerged spintronic-based neuromorphic computing.

### ACKNOWLEDGMENTS

This work is part of the Gravitation program "Research Centre for Integrated Nanophotonics," which is financed by the Netherlands Organisation for Scientific Research (NWO). This project has also received funding from the European Union's Horizon 2020 research and innovation program under the Marie Sklodowska-Curie (Grant Agreement No. 860060).

## AUTHOR DECLARATIONS

### Conflict of Interest

The authors have no conflicts to disclose.

### Author Contributions

**H. Pezeshki:** Conceptualization (lead); Data curation (lead); Formal analysis (lead); Investigation (lead); Methodology (lead); Software (lead); Validation (lead); Visualization (lead); Writing – original draft (lead); Writing – review & editing (lead). **P. Li:** Visualization (supporting); Writing – original draft (supporting); Writing – review & editing (supporting). **R. Lavrijsen:** Project administration (supporting); Resources (supporting); Supervision (supporting); Validation (supporting); Writing – review & editing (supporting). **M. Heck:** Methodology (supporting); Writing – review & editing (supporting). **B. Koopmans:** Conceptualization (supporting); Formal analysis (supporting); Funding acquisition (equal); Methodology (equal); Project administration (lead); Resources (lead); Supervision (lead); Writing – review & editing (equal).

## DATA AVAILABILITY

The data that support the findings of this study are available from the corresponding author upon reasonable request.

12 October 2024 11:14:55